\documentclass[aps,physrev,twocolumn,groupedaddress]{revtex4-2}
\usepackage{graphicx}
\usepackage{amsmath}
\usepackage{url}         

\begin{document}


\title{Quantum Interference Driven Electron Transport through Fano-Anderson Systems}


\author{Koushik R. Das, Sudipta Dutta}
\email[]{sdutta@labs.iisertirupati.ac.in}
\affiliation{Indian Institute of Science Education and Research (IISER) Tirupati, Tirupati, Andhra Pradesh, India}


\date{\today}

\begin{abstract}
   We investigate the electronic transport behavior of Fano-Anderson (FA) systems, consisting of a one-dimensional finite backbone chain and an attached side-group of varying length. The tight-binding model within the non-equilibrium Green's function (NEGF) reveals distinct interference patterns between the frontier orbitals of odd and even side-group systems that lead to distinct transport mechanisms across various configurations. The even side-group systems exhibit resonant tunneling peaks, while the odd ones demonstrate significant destructive interference at Fermi energy leading to antiresonance and reduced current responses. Further exploration of the role of electron-electron interactions within mean-field Hubbard Hamiltonian indicates the emergence of negative differential resistance (NDR) in odd side-group systems at lower biases, a phenomenon absent in the non-interacting case. Our study demonstrates the critical role of quantum interferences among the frontier molecular orbitals and electron correlations in shaping the transport behavior of FA systems, providing insights for the design of molecular-scale electronic devices. 
\end{abstract}


\maketitle

\section{Introduction}
   Electron conduction through a single molecule has been of sustained interest from both fundamental and application perspective, owing to the significant potential to revolutionize the design and functionality of nanoscale devices\cite{huang2020recent,gupta2023nanoscale}. At core of this discipline lies the intricate interplay between molecular structure and electronic transport properties, which can be tuned through various parameters, like relative electrode orientation, structural modifications, attaching functional groups and so on\cite{metzger2018quo}. Among the diverse models employed to understand these phenomena, the Fano-Anderson (FA) system stands out due to its unique ability to exhibit interesting interference effects arising from the coupling of discrete molecular states with the electrode continuum\cite{nozaki2012control,pan2023designing}.
   
   Fano-Anderson systems are described by a linear array of coupled sites with one or more defects connected locally with various geometries \cite{mahan2013many}. One of such systems is depicted in Fig.1. These systems have been studied extensively to probe quantum waveguides \cite{aligia2004magnetotransport,gores2000fano,fan2002sharp} such as microring resonators \cite{pereira2002gap} and photonic crystal waveguides \cite{miroshnichenko2005sharp}. Various methods such as decimation renormalization method\cite{chakrabarti2007fano} and Frenkel-Kontorova model\cite{miroshnichenko2005engineering} have been employed to understand the transport mechanism through such systems. In this study, we investigate the electron transport characteristics of Fano-Anderson systems with varying side-group lengths, focusing on the role of electron-electron interactions and their impact on transport behavior. We model the system within tight-binding Hamiltonian with further inclusion of onsite Coulomb correlation term using the mean-field Hubbard Hamiltonian which allows us to explore the effects of $e-e$ interactions on the transport behavior of the said systems. Our analysis reveals that the interference between the highest occupied molecular orbitals (HOMO) significantly influence the transmission properties, leading to distinct behaviors in the systems with odd and even side-group configurations.
   
   We present a comprehensive examination of electron transport though such systems, highlighting the emergence of Fano resonances and their implications for negative differential resistance (NDR) under applied bias. By employing the analyses of frontier orbitals and their quantum interference, we elucidate the mechanisms underlying these transport phenomena, thereby providing insights into the fundamental principles governing electron transport in molecular devices.
   \vspace{-0.67cm}

\section{Theoretical Details}
   Our model comprises of a linear chain (backbone) having seven sites (termed as 7c) and a side-group of varying length, termed as $n$s, with n being the number of sites. The side-group is attached at the middle site of the 7c backbone, as shown in Fig.1.
   
   \begin{figure}[h]\centering
   	\includegraphics[width=0.4\textwidth]{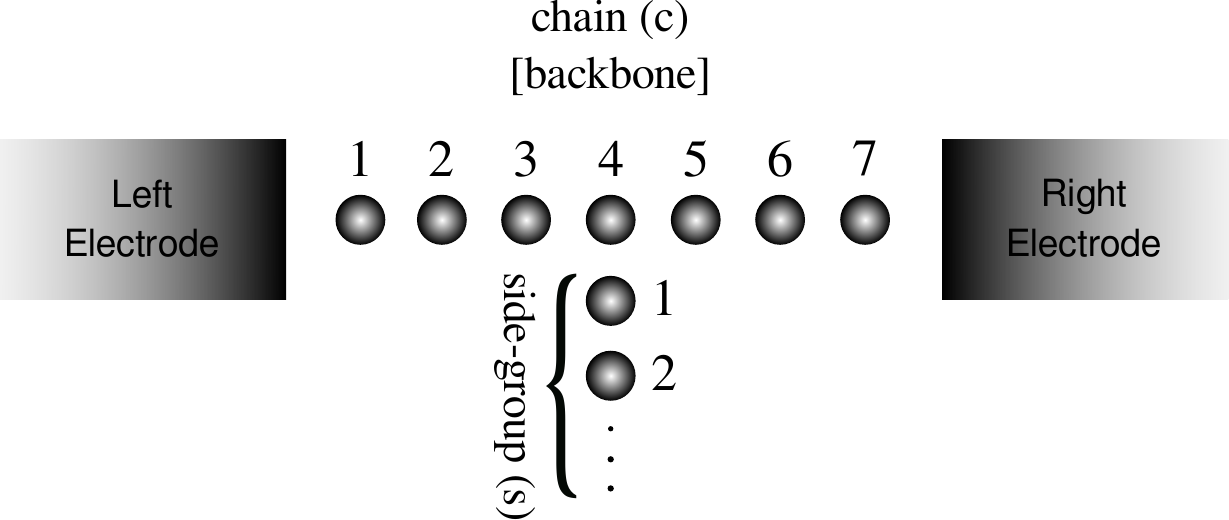}
   	\caption{Schematic illustration of a Fano-Anderson (FA) system with 7 site backbone chain with a side group attachment of varying length. Electrodes are attached at sites 1 and 7 of the backbone chain for transport calculations.}
   \end{figure}
   
   The Hubbard Hamiltonian for the molecular system under the external bias is expressed as:
   \begin{equation}
   	 \begin{split}
   	H_{mol} &= \sum_{\langle i,j \rangle \sigma} t(a_{i,\sigma}^\dagger a_{j,\sigma}+a_{j,\sigma}^\dagger a_{i,\sigma}) \\& + \sum_{i}U n_{i\uparrow}n_{i\downarrow} - \sum_{i,\sigma}V_i a_{i,\sigma}^\dagger a_{i,\sigma}
	 \end{split}
   \end{equation}
   \vspace{-0.4cm}
   
   \noindent where $a_{i,\sigma}^\dagger$ ($a_{i,\sigma}$) creates (annihilates) an electron with spin $\sigma$ at $i$-th site. The number operators are defined as $n_{i,\sigma} = a_{i,\sigma}^\dagger a_{i,\sigma}$. The parameters $t$ and $U$ represent the nearest neighbor hopping integral and the onsite Coulomb potential, respectively. The term $V_i$ is  the site potential, which originates from the applied external bias simulated as a ramp, and can be formulated as \cite{das2024asymmetric}:
   
   \begin{equation}
   	V_i=-\frac{V_L r_{iL} + V_R r_{iR}}{r_{iL}+r_{iR}}
   \end{equation}
   
   \noindent The term $V_{L/R}$ represents the potential at the corresponding contact-site, attached to the left ($L$) or right ($R$) electrode, with $r_{i(L/R)}$ denoting the vectorial distance between the $i^{th}$ site and the respective electrode. The value of $V_{L/R}$ is assumed to be $\pm V/2$, where $V$ represents the potential difference between the left and right electrodes. We consider $t$ as the unit of energy for our calculations. 
   
   Using mean-field approximation\cite{bruus2004many} for weakly correlated systems, the many-particle Hubbard interaction term can be approximated within a single-particle picture as:
   
   \begin{equation}
   	\sum_{i}U n_{i\uparrow}n_{i\downarrow} \approx \sum_{i}U [ n_{i\uparrow} \langle n_{i\downarrow} \rangle + n_{i\downarrow} \langle n_{i\uparrow} \rangle - \langle n_{i\uparrow} \rangle \langle n_{i\downarrow} \rangle ]
   \end{equation}
   
   The quantities within the corner brackets are the average densities. We consider the anti-ferromagnetic state with alternating up and down spin densities along the 7c backbone and the $n$s side-group as the initial guess, where the first site on the side-group in coupled ferromagnetically with the middle site of the backbone. Starting from this initial guess density, we then self-consistently solved the mean-field Hubbard Hamiltonian. Subsequently, we integrate this with the Non-Equilibrium Green's Function (NEGF) formalism to compute the transmission and current as a function of the applied bias.
   
   The current is determined by integrating the transmission function $T(E,V)$, following the Landauer-B\"{u}ttiker formula\cite{datta1997electronic}:

   \begin{equation}
   	I(V)= {\frac{2e^{2}}{h}} \int_{\mu_{L}}^{\mu_{R}} T(E,V) dE 
   \end{equation}
   
   \noindent where $e$ represents the electronic charge, $h$ denotes Planck's constant, $V$ stands for the applied bias, and $E$ represents the energy of the electron. $\large\mu\small_{L/R}$  are the chemical potentials of the $L/R$ electrodes, defined as $\large\mu\small_{L/R} = E_F \mp eV/2 $. The Fermi energy of the electrode at zero-bias is denoted by $E_F$. The interval $[\mu_L(V),\mu_R(V)]$, or equivalently $[-V/2,V/2]$, signifies the energy range contributing to the current integral, known as the bias window. This conforms to the understanding that electrons near the $E_F$ will primarily contribute to the total current.
   
   The transmission function $T(E,V)$ represents the overall probability of an electron with energy $E$ to traverse between electrodes through the device region (see Fig. 1), and it can be mathematically defined as\cite{landauer1992conductance,buttiker1986four,datta1997electronic}:
   
   \begin{equation}
   	T(E,V)=Tr[\Gamma_L(E,V)G(E,V)\Gamma_R(E,V)G^\dagger(E,V)]
   \end{equation} 
   
   \noindent In this expression, $G(E,V)$ is the retarded Green's function:
   \begin{equation}
   	\begin{split}
   	G(E,V)&=[(E+i\eta)I-H]^{-1}\\&=[(E+i\eta)I-H_{mol}-\Sigma_L-\Sigma_R]^{-1}
   	\end{split}
   \end{equation}
   \noindent where the device Hamiltonian is given by $H$, composed of the molecular Hamiltonian $H_{mol}$ and self energies $\Sigma_{L/R}$, that appear due to the coupling of the molecule with $L/R$ electrodes. $I$ is the identity matrix and $\eta$ is an infinitesimally small number to avoid numerical divergence. The expression $\Gamma_{L/R}$, termed as the spectral density, characterizes the broadening of eigenstates attributed to the $L/R$ electrodes. This broadening arises from the coupling of the molecule with the respective electrodes and is mathematically defined in terms of the self-energies $\Sigma_{L/R}$. The spectral densities can be written as:

   \begin{equation}
   	\Gamma_{L/R}=i~ (\Sigma_{L/R}-\Sigma^\dagger_{L/R})
   \end{equation}
   
   \noindent We set $\Sigma_{L/R}=0.05$. The external bias is systematically varied from $-1V$ to $+1V$ in increments of $0.1V$. For each bias voltage, we compute the corresponding transmission function and current. 
\section{Results and Discussions}
   We investigate the electron transport through the FA systems with varying side-group length. As the systems (in Fig.1) and the electrode contacts are structurally symmetric, the $I-V$ characteristics are completely symmetric across the positive and negative bias regimes. We shall first analyze the results for the tight-biding (TB) regime i.e. $U=0$, and then incorporate the e-e interactions ($U \ne0$) within mean-field regime to understand their effect on the conduction.
   
   \subsection{Tight-binding (TB) limit (U=0) }
       The $I-V$ characteristics in the positive bias regime are presented in the Fig.2(a). Clear distinction between the even and odd side-group length systems can be observed. For even side-group lengths ($2s$, $4s$ etc., including the system with no side-group), the current shows a steep increase at a lower bias. However, the current responses among this class of side-groups do not show much variation until at high bias. On the contrary, for odd side-group lengths ($1s$, $3s$ etc.) the current response is much lower as compared to the previous case and there is a clear distinction between the different chain lengths, as the current response increases with the chain length in this class of side-groups.  
       
       \begin{figure}[h]\centering
       	\includegraphics[width=0.4\textwidth]{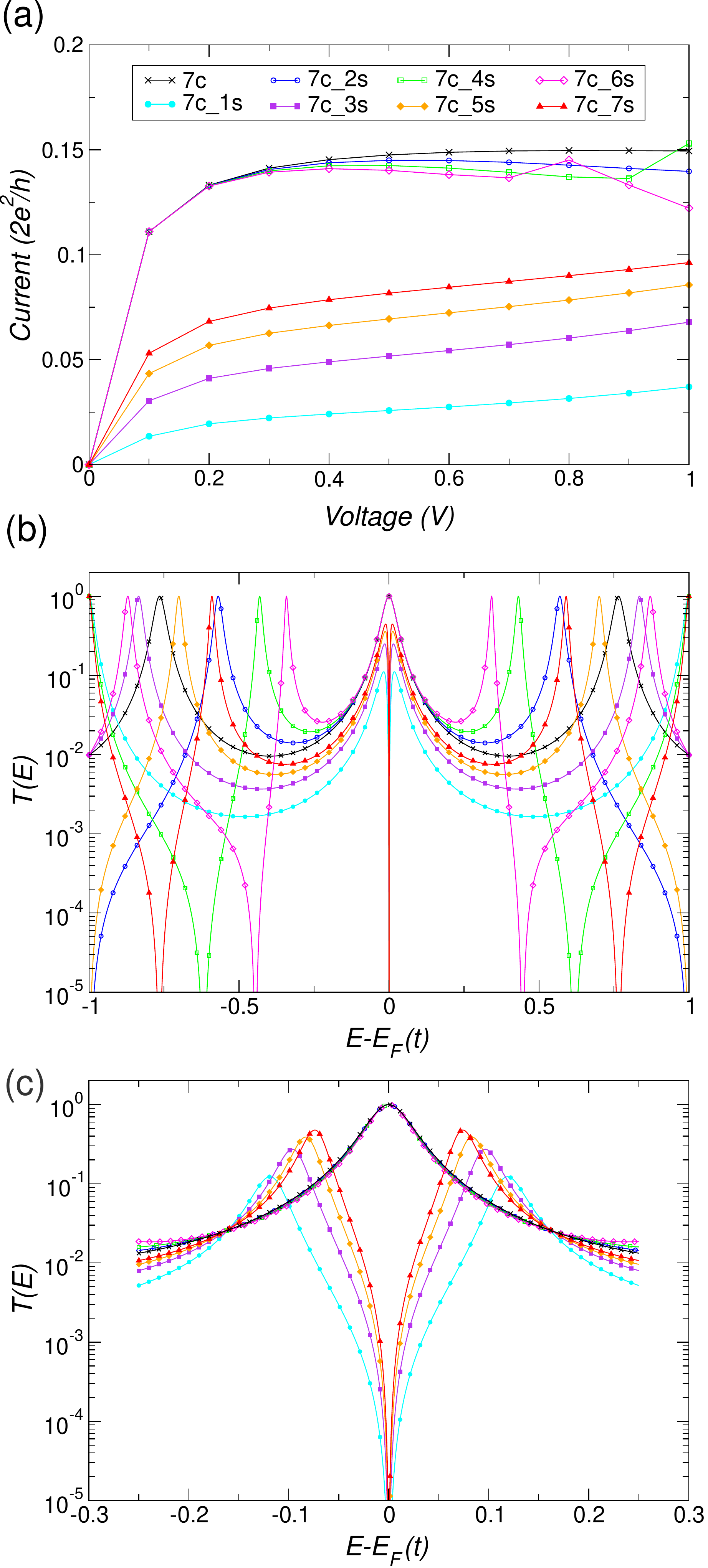}
       	\caption{(a) The $I-V$ characteristics and normalized transmission functions at (b) zero-bias and (c) $0.5V$ for the FA systems with non-interacting ($U=0$) limit.}
       \end{figure}
       
       This distinction between the odd and even side-groups can be effectively understood from the zero-bias transmission functions as shown in Fig.2(b). It can be seen that all the odd side-group systems exhibit an antiresonance dip at the $E_F$, whereas the even side-group systems has a distinct peak. These respective features are found to be robust under application of bias, as can be seen in Fig.2(c). Such difference in the transmission with odd and even side-group length have been reported previously\cite{miroshnichenko2005engineering,chakrabarti2007fano}. A notable feature common among the systems irrespective of odd/even side-group lengths are the Fano resonances. It comprises a Breit-Wigner peak due to resonance with an eigenvalue of the system and an antiresonance arising from destructive interference between MOs within a narrow energy range\cite{lambert2015basic}. As can be seen from Fig.2(a) and (b), the Fano resonance leads to NDR behavior for 7c-6s system. Since the Fano resonance falls within the bias window for this system, the NDR is visible within the reported bias window till $1V$. However, the other systems also exhibit the similar behavior beyond this bias window due to prominent Fano feature further from the $E_F$. However, at larger bias, the non-frontier MOs start contributing towards the net current and that masks the observed odd-even effect in the current response.
       
       To understand the features of transmission functions, orbital analysis of the conduction channels i.e. MOs of the systems play an important role\cite{stokbro2003theoretical,xue2001charge,li2012orbital}. In this regard, the orbital symmetry rule proposed by Yoshizawa et.al \cite{yoshizawa2008orbital} has been considered to be a useful technique. Considering weak molecule-electrode coupling, the zeroth-order Green's function is given by:
       \begin{equation}
       	G^0_{ab}(E)=\sum_{k} \frac{C^*_{ak}C_{bk}}{E-\varepsilon_k \pm i\eta}
       \end{equation}
       
       \noindent where, $C_{ak}$ and $\varepsilon_k$ are the coefficient at terminal site-$a$ and eigenenergy for the $k^{th}$ MO, respectively. Here the Green's function can be considered as retarded or advanced based on the sign of the $\eta$ term, which is an infinitesimal number used to avoid the zero-division divergence. The frontier molecular approximation states that only the frontier orbitals i.e. the HOMO and LUMO are the major contributors towards the electronic properties in $\pi$-conjugated systems. Therefore, the summation in the Green's function in Eqn.(8) can be approximated to the limits of only HOMO and LUMO. However, in case of alternate hydrocarbon systems, as considered here, the signs of the AO coefficients are opposite for MOs having particle-hole symmetry, i.e., the MOs energetically equidistant from the $E_F$. This is depicted in Fig.3 for a few representative systems. As a result, there contributions towards the zeroth-order Green's function cancel each other. In the present case, however, the systems are  having nonbonding molecular orbitals (NBMOs) that are singly occupied and appears exactly at the $E_F$, as shown in Fig.3. Therefore, the summation in Eqn.(8) can be restricted only to the NBMOs. This approach has been successful in explaining the conductance in a variety of systems \cite{okazawa2023frontier,tsuji2011orbital,tsuji2017frontier,tsuji2012orbital}. Therefore, we adopt similar analysis for the FA systems here. Note that NBMOs are having two-fold degeneracy for odd side-group systems and non-degenerate for even side-group systems, as shown in Fig.3.
       
       \begin{figure}[h]\centering
       	\includegraphics[width=0.5\textwidth]{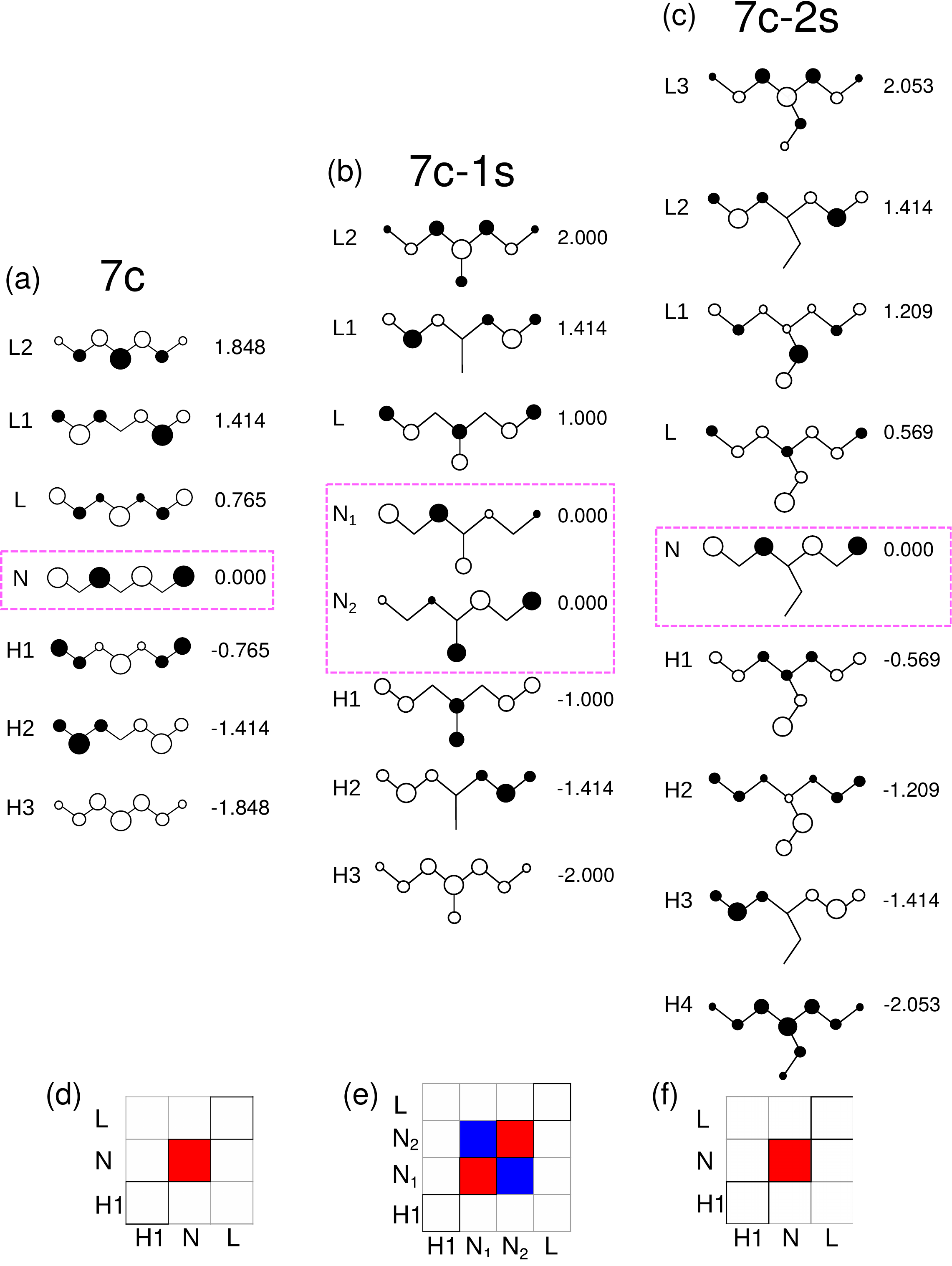}
       	\caption{The MOs for (a) 7c (b) 7c-1s (c) 7c-2s systems at the Fermi energy. The hollow (filled) circles imply negative (positive) atomic orbital coefficients, while the size of the circles are proportional to their magnitudes. The corresponding Q-matrices at $E_F$ are shown in (d,e,f), respectively.}
       \end{figure}
       
       Since the electrodes are attached at the terminal ends of the backbone chain (see Fig.1), the coefficients in Eqn.(8) are $C_{1k}$ and $C_{7k}$. Using the NBMO approximation, described above, the zeroth-order Green's function evaluated at $E_F$, for odd side-group systems would be:
       
   
        \begin{equation}
        	G^0_{1,7}(E_F)=\frac{C^*_{1,NBMO_1}C_{7,NBMO_1}}{E_F-\varepsilon_{NBMO_1}\pm i\eta} ~+  \frac{C^*_{1,NBMO_2}C_{7,NBMO_2}}{E_F-\varepsilon_{NBMO_2}\pm i\eta}
        \end{equation}
       
       For the even side-group systems, Eqn.(9) contains only one term, as there is only a single NBMO in such cases. It can be easily seen from the Fig.3 that the product $C^*_{1,NBMO_i}C_{7,NBMO_i}$ is always negative due to opposite signs of coefficients in two terminals. Hence their sum, both for odd and even side-group cases is always non-zero. This, in turn, implies that the Green's function in Eqn.(9) is non-zero and hence the transmission can never exhibit antiresonance at $E_F$, which is contrary to the results obtained using the NEGF calculations (see Fig.2(b)). Therefore, we observe that the orbital symmetry rule is applicable for even side-groups only and it fails to explain the transmission behavior of the odd side-group systems. It demands an alternative approach to explain the role of MOs in the transmission functions of the FA systems.

       It has been observed in the reported literature that the antiresonance in transmission happens due to the destructive quantum interference (QI) between the frontier orbitals\cite{das2024asymmetric,yoshizawa2008orbital,gunasekaran2020}. It can be effectively tracked and visualized using the Q-matrix formalism proposed by Gunasekaran et.al.\cite{gunasekaran2020}. Under the NEGF methodology, the Hamiltonian and the Green's functions are expressed in the AO basis. Extraction of the QI information necessitates a basis transformation from the AO to MO basis, which can be achieved through a matrix P containing the eigenvectors of G as columns, such that $P^{-1}GP$ is a diagonal matrix. After the transformation to the MO basis, the resulting transmission function T appears as a sum over all the MOs, comprising both the interfering off-diagonal terms ($T_{ij}$) and diagonal terms ($T_i$):
       \vspace{-0.4cm}
       
       \begin{equation}
       	T=\sum_{i} {T_i} + \sum_{ij} {T_{ij}}
       \end{equation}
       This is obtained  by evaluating the Q-matrix:
       \begin{equation}
       	Q(E)=(P^\dagger \Gamma_L G P) \circ (P^{-1} \Gamma_R G^\dagger {{P^{-1}}^\dagger})^T
       \end{equation}
       Here, $[\circ]$ denotes the entrywise (Schur) product, and $[*]^T$ implies matrix transpose. It can be easily shown that,
       \vspace{-0.4cm}
       
       \begin{equation}
       	T=\sum_{ij} Q_{ij}
       \end{equation}
       \vspace{-0.2cm}
       
       This expression indicates that the Q-matrix and the transmission function $T$ are equivalent. We use this formalism to gain insight into the QI effects in the transmissions by calculating the Q-matrices for specific energies and bias for the FA system and represented in the form of color-maps, where blue and red colors indicate destructive and constructive interferences, respectively, with the color intensity indicating the magnitude of the interference.  Note that the diagonal is aligned bottom-left to top-right. The color-maps of Q-matrices for the FA systems at Fermi energy under zero-bias is presented in Fig.2(c).
       
       The transport at Fermi energy is NBMO-dominated for all the systems, as can be seen from Fig.4. The difference lies in the interference patterns of the odd and even side-group systems. The even side-group systems have contribution from the NBMO only, which leads to the resonant tunneling peak in transmission at $E_F$. On the other hand, the odd side-group systems have two-fold degenerate NBMO levels, which exhibit strong destructive interference with each other and nullify their individual contributions, leading to a prominent antiresonance in transmission at $E_F$ and an overall lower current response as compared to the even side-group systems (see Fig.2(b)).  From this analysis, it is evident that the interference behavior among the MOs are directly responsible for their observed transmissions and are the main reason for the  distinction between odd/even side-group FA systems.
       
       To further gain insight about the transport and quantum interference behavior in presence of the $e-e$ interactions, which is very crucial in such quantum confined system, we model the FA systems within Hubbard Hamiltonian and solve the same within the mean-field approximation.
       \vspace{-0.4cm}
       
   \subsection{Mean-field limit (U$\ne$0) }
       We consider the $e-e$ interaction in the mean-field limit (Eqn.3), which converts the many-particle Hubbard term into an effective single-particle term. This enables us to combine it with the NEGF and calculate the transport properties of the FA systems at half-filling. 
       
       \begin{figure}[h]\centering
       	\includegraphics[width=0.5\textwidth]{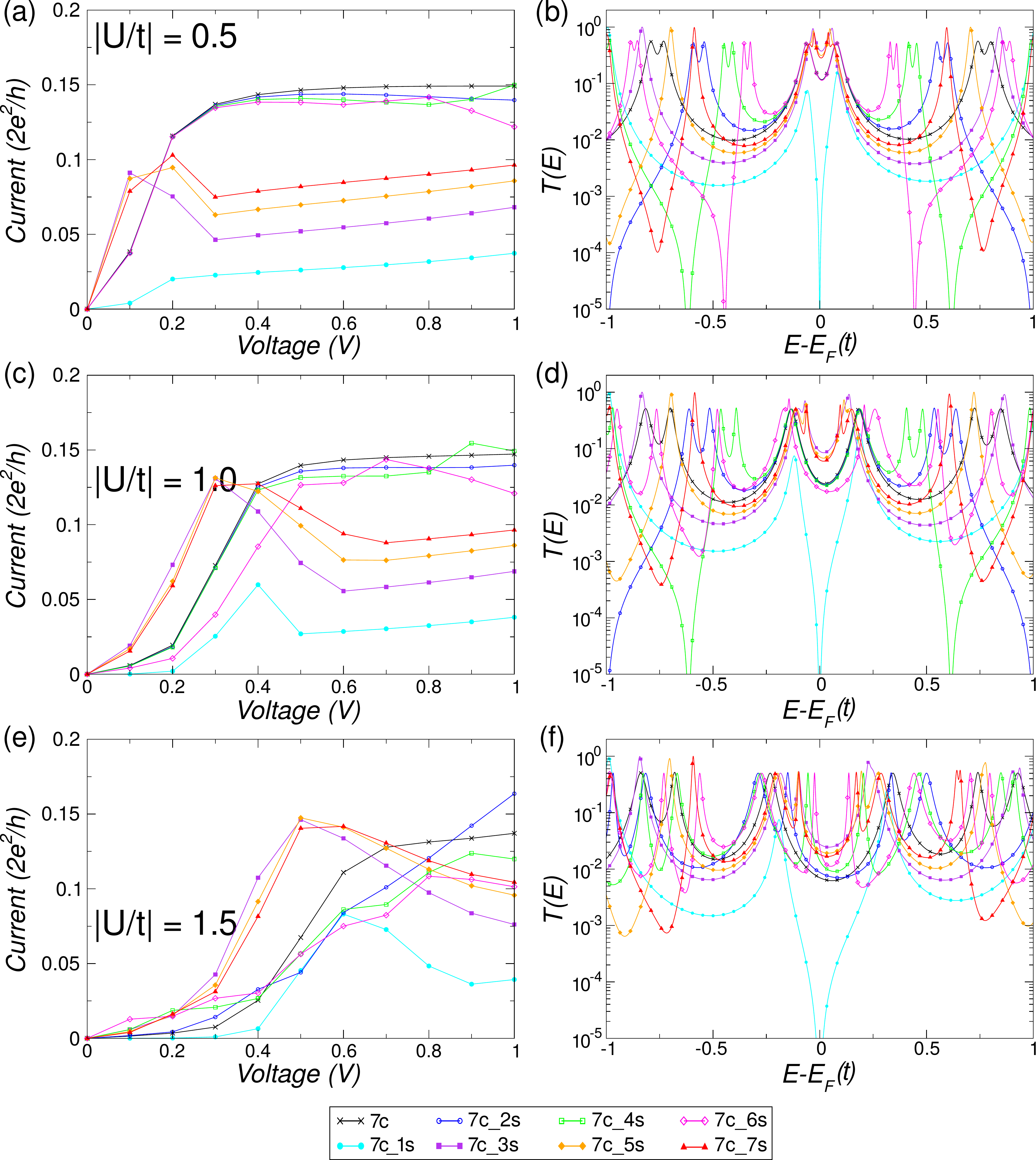}
       	\caption{The $I-V$ characteristics and zero-bias normalized transmission functions calculated in the mean-field limit with half-filling for $|U/t|=0.5$ (a, b), $|U/t|=1.0$ (c, d) and $|U/t|=1.5$ (e, f), respectively.}
       \end{figure}
   
       The I-V characteristics for different $e-e$ interaction strengths for all the systems are shown in Fig.4 along with the corresponding zero bias transmission plots. For lower $e-e$ interaction strength of $|U/t|=0.5$, the distinct current responses for odd and even side-group systems are observed, resembling the tight-binding results. However, with high interaction strength, the distinction between different systems become ambiguous. 
       
       Interestingly, the inclusion of the $e-e$ interaction eliminates the destructive interferences between the frontier orbitals at $E_F$ and results in the disappearance of the antiresonance in the transmission plots for odd side-group systems, that is observed in the tight-binding results. The only exception is the 7c-1s system, that still features the antiresonance behavior at $E_F$. This behavior gets widened with increase in $e-e$ interaction strength. Application of small bias makes this antiresonance disappear. However, gradual increase of bias in all odd side-group systems reinforces the destructive interferences and the consequent antiresonance at the $E_F$, beyond a certain bias that depends on the system and the $e-e$ interaction strength. Such observations indicate a competition between the $e-e$ interaction strength that tries to localize the electrons and the external bias that drives the electrons in the system. This can be further established by the delayed onset of the current with rising bias as the $U/t$ ratio increases.
       
       Further inspection of the transmission functions reveals the splitting of the $U=0$ peaks (in Fig.2(b)) in the $U\ne0$ cases (Figures 4(b), (d) and (f)) for the even side-group systems. This phenomena indicates breaking of spin-symmetry. Note that, we are considering the mean-field Hubbard Hamiltonian here at half-filling with explicit consideration of the spin orbitals. Since, the even side-group systems are having overall odd number of atoms, owing to the odd number of atoms in the backbone (7c), at half-filling, there is asymmetry between the number of up and down spins. This breaks the spin symmetry. As a result, each peak in the $U=0$ transmission plot gets splitted into two peaks. However, the spin-symmetry in odd side-group systems keeps the transmission peaks intact.
       
       \begin{figure}[h]\centering
       	\includegraphics[width=0.45\textwidth]{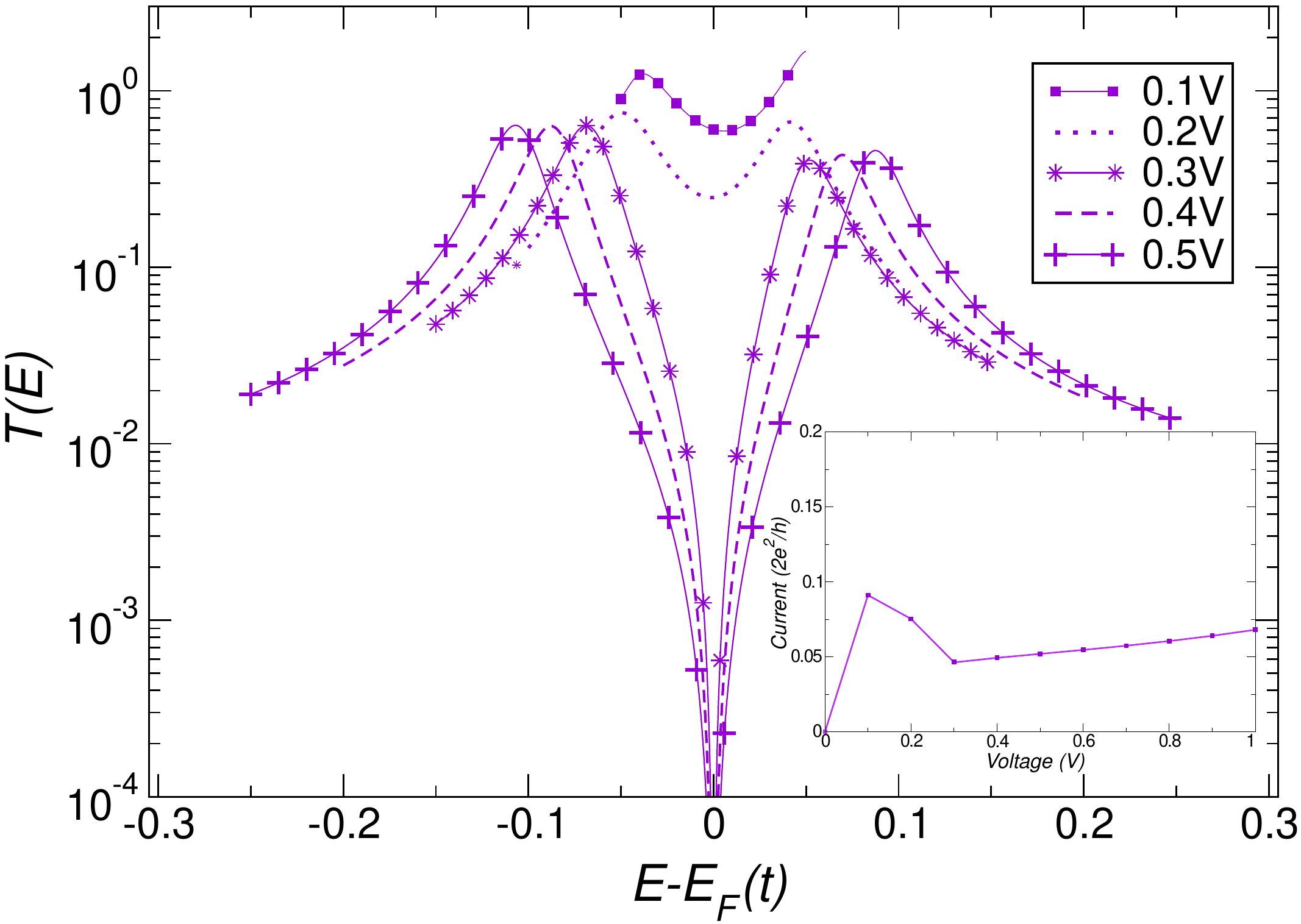}
       	\caption{Transmission functions within the bias window $[-V/2,V/2]$ at a given bias values for the 7c-3s system with $|U/t|=0.5$. Inset shows the corresponding $I-V$ characteristics.}
       \end{figure}    
       
       \begin{figure}[h]\centering
       	\includegraphics[width=0.3\textwidth]{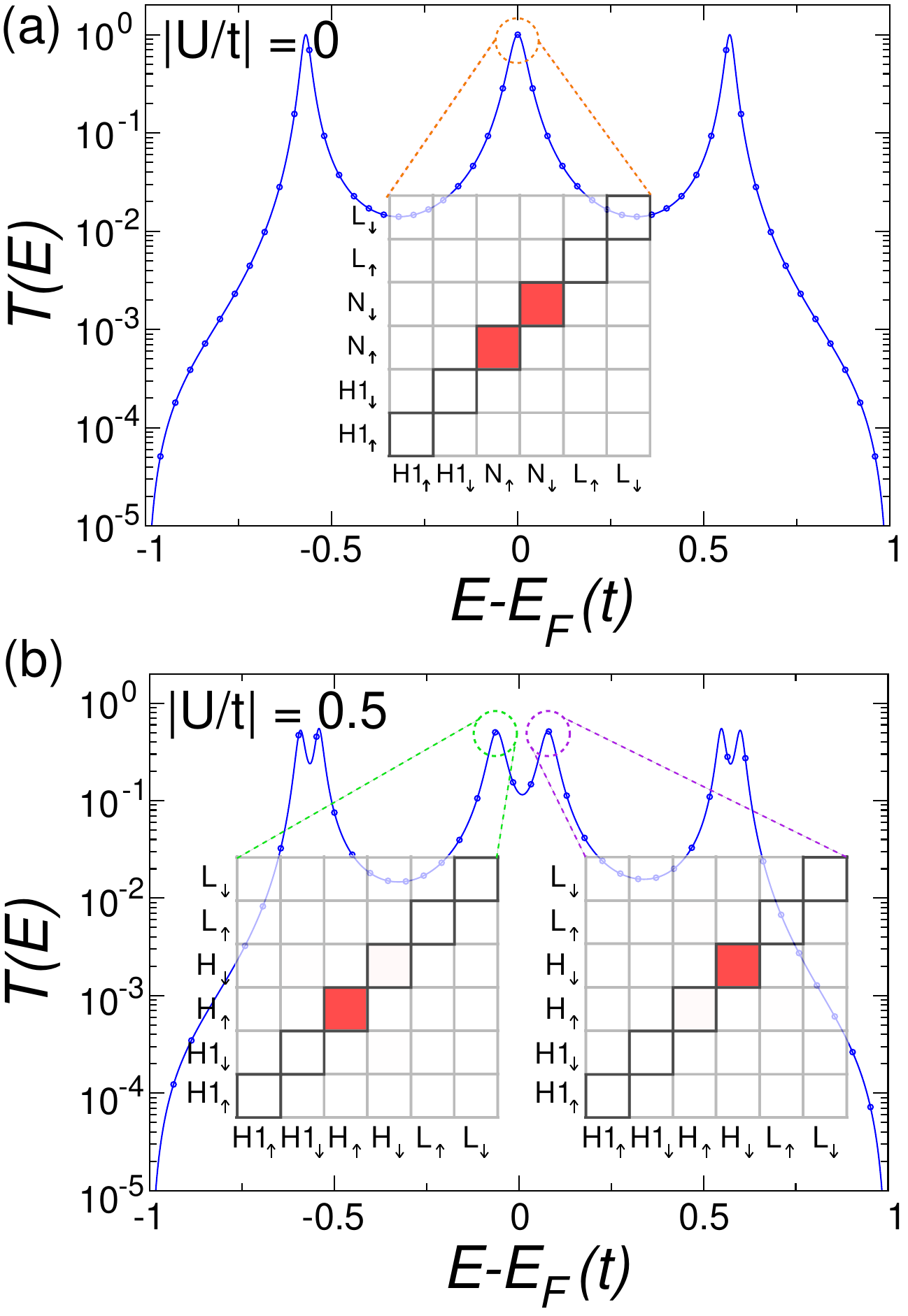}
       	\caption{Zero-bias normalized transmission functions for the 7c-2s system with (a) $|U/t|=0$ and (b) $|U/t|=0.5$. The respective Q-matrices at transmission peaks (circled) near the $E_F$ are provided as insets.}
       \end{figure}
       
       These discussion can be substantiated from the perspective of interferences between spin-resolved MOs and are presented in Fig.6. The spin-resolved and unresolved (spatial) Q-matrices are due to spin-resolved and spatial MOs respectively, and are related simply as:
       
       \begin{equation}
       	Q_{ij}=Q_{ij}^\uparrow + Q_{ij}^\downarrow
       \end{equation}
       
       As an example for odd electron systems, we look into the zero-bias transmission and corresponding Q-matrices for the 7c-2s system. In the tight-binding picture, the zero-bias transmission peak at $E_F$ appears due to the constructive contributions from orbitals (NBMO)$_\uparrow$ and (NBMO)$_\downarrow$ (see Fig.6(a)). This peak splits into two due to the presence of $e-e$ interactions (Fig.6(b)) which breaks the spin-degeneracy discussed above, and each peak is contributed by a single spin-MO. The peaks at other energies can also be explained similarly. 
       
       \begin{figure}[h]\centering
       	\includegraphics[width=0.3\textwidth]{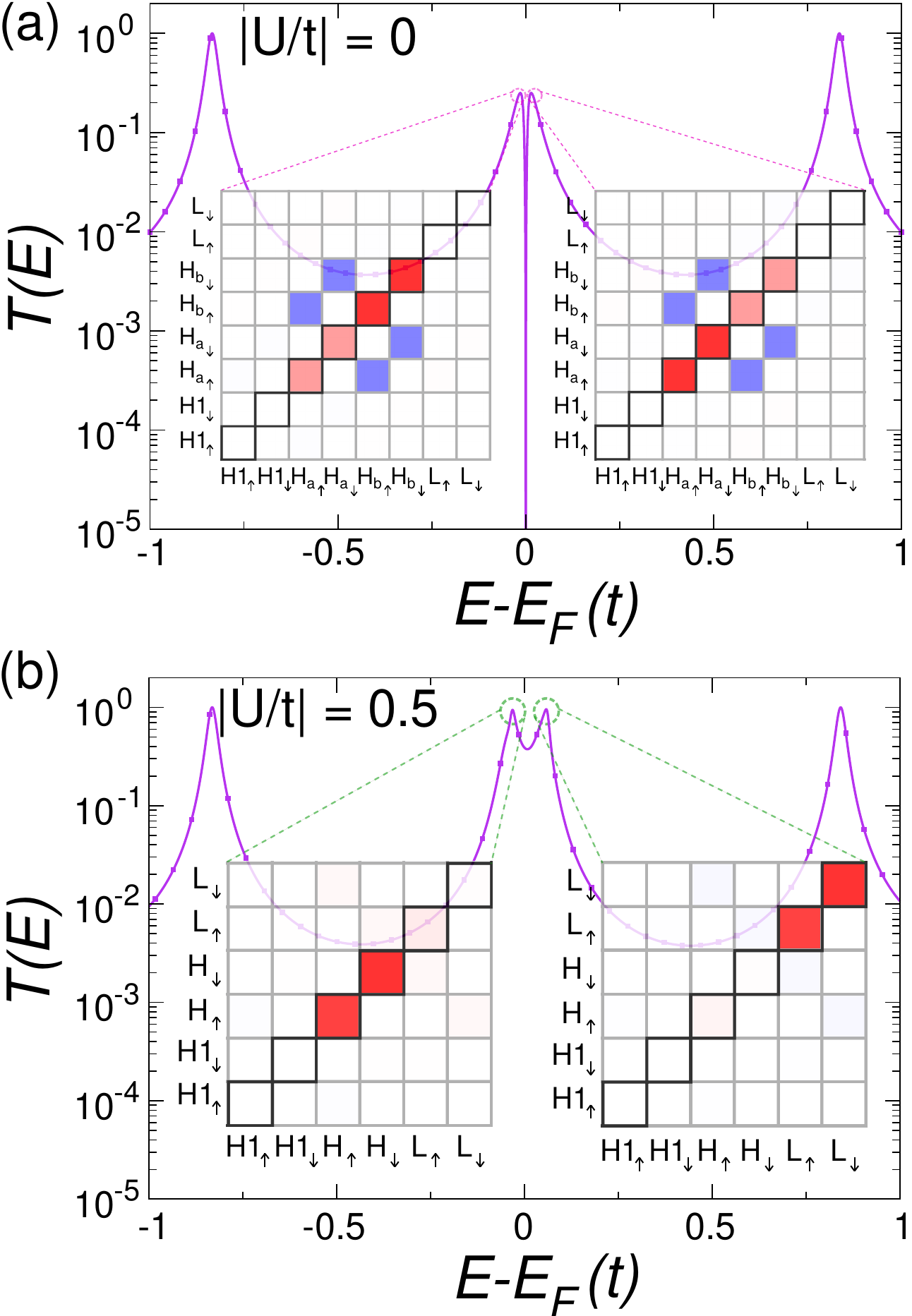}
       	\caption{Zero-bias normalized transmission functions for the 7c-3s system with (a) $|U/t|=0$ and (b) $|U/t|=0.5$. The respective Q-matrices at transmission peaks (circled) near the $E_F$ are provided as insets.}
       \end{figure}
       
       For comparison with even $N_e$ systems, we take the 7c-3s as an example. Q-matrices at $-0.83t$ are presented corresponding to the peak at this particular energy point in the transmission, both for the TB and mean-field limits (Fig.7). Here the peak-splitting do not occur for non-zero $U$ since $N_\uparrow = N_\downarrow$ in this case, and this holds for odd side-group systems in general. 
       
       A notable observation is the appearance of NDR at lower biases for odd side-group systems. This feature is absent for $U=0$ case (Fig.2(a)). To get further insight of this NDR behavior, we investigate the bias dependent transmission behavior and present the same for the 7c-3s with $|U/t|=0.5$ in Fig.5. As has been mentioned before, the antiresonance at the $E_F$ vanishes after the inclusion of the $e-e$ interaction. This leads to the initial increase in the current response on application of bias. At higher bias, the appearance of destructive interferences between the frontier orbitals brings back the antiresonance at $E_F$. Moreover, with increasing bias strength, the antiresonance widens and pushes away the resonant peaks away from the bias window, resulting in a slowly rising current. Incorporation of these resonant tunneling peaks at higher biases beyond $1V$ can lead to a jump in the current response.
       
       \vspace{-0.5cm}
       
\section{Conclusion}  
    To summerize, we investigate the electron transport through a particular class of FA systems, containing a fixed backbone chain and varying side-group length. Using the TB method coupled with the NEGF approach, we calculate the corresponding transmissions and current responses through the systems with the odd and even side-group lengths. Orbital symmetry rule is applied to explain these findings, but has been found to be insufficient. Instead, implementation of the Q-matrix formalism gives an in-depth understanding of the interference behavior among MOs. The destructive interferences between the degenerated HOMO levels lead to the transmission antiresonance at Fermi energy in the odd side-group systems, which is absent in case of the even side-group systems. Inclusion of $e-e$ interactions within mean-field Hubbard Hamiltonian enables us to explore the effect of spin symmetry in the transport through these systems. For low interaction strength, the odd and even side-groups can be distinguished in terms of their current responses, which gets obscured at higher interaction strengths. Odd side-group systems exhibit NDR behavior, which can be explained from the reappearance of the transmission antiresonance at higher biases within the bias window. Another interesting feature is the splitting of the transmission resonance peaks, which occurs only for the even side-group systems due to spin symmetry breaking. This is validated from the analysis of the spin-resolved Q-matrices, which show that each of the splitted peaks is due to a single MO of a particular spin type. This feature is absent for the odd-side group systems since they contain equal number of spins of both types. It would be interesting to explore the systems away from half-filling i.e., with electron or hole-doping cases in such FA systems. This can open up the possibility of spin-filtering effect in FA systems.
	\vspace{-0.5cm}

%



\begin{acknowledgments}
   KRD and SD thank IISER Tirupati for intramural funding and the Science and Engineering Research Board (SERB), Department of Science and Technology (DST), Government of India for research grant CRG/2021/001731.
\end{acknowledgments}

%

\end{document}